\documentclass[aps,prb,twocolumn,groupedaddress,floats]{revtex4}

\usepackage{graphicx}
\usepackage{dcolumn}
\usepackage{amsmath}

\begin{document}



\title{Strong-coupling scenario of a metamagnetic transition}




\author{D.\ Meyer}
\email[]{d.meyer@ic.ac.uk}
\affiliation{Department of Mathematics, Imperial College,
  180 Queen's Gate, London SW7 2BZ, United Kingdom}
\author{W.\ Nolting}
\affiliation{Lehrstuhl Festk{\"o}rpertheorie, Institut f{\"u}r Physik,
  Humboldt-Universit{\"a}t zu Berlin, Invalidenstr.\ 110, 10115 Berlin, Germany}

\date{\today}

\begin{abstract}
We investigate the periodic Anderson model in the presence of an
external magnetic field, using 
dynamical mean-field theory in combination with the modified
perturbation theory. A metamagnetic transition is observed which
exhibits a massive change in the electronic properties. These are
discussed in terms of the quasiparticle weight and densities of
states. The results are compared with the experimental results of the
metamagnetic transition in CeRu$_2$Si$_2$.
\end{abstract}


\maketitle

By ``metamagnetic transition'', we describe an anomalous behavior of
the magnetization as function of external field $b_{\rm ext}$, namely a
sudden increase at a finite field $b_{\rm ext}^*$.
Many rare-earth materials exhibit this kind of behavior. One
has to distinguish between those materials which already show
long-range (e.g.\ antiferromagnetic) order for zero-field and those
which are paramagnetic. 

In the first case, the materials already have finite local moments at
zero-field. Here a metamagnetic transition occurs when the external
field is stronger than the internal (antiferromagnetic) exchange
between these moments. This situation is found for example in
CeFe$_2$-based alloys\cite{Mea00}. 

The other possibility, where a paramagnetic substance enters a
high-magnetization state at a critical field $b_{\rm ext}^*$  is
realized by some heavy-fermion metals as
for example CeRu$_2$Si$_2$\cite{Mea91,AUAO93,Sea95,Aea98} or
UCoAl\cite{Aea98,MGKYASIS99}. 
As will be discussed further below, experiments indicate a substantial
change in the electronic structure at the transition, which is not yet fully
understood. 

In this paper, we investigate a similar
transition found in a relatively simple
electronic model of heavy-fermion compounds.
We examine the periodic Anderson
model (PAM) with an external magnetic field. The periodic Anderson model
describes the interplay between strongly correlated localized electrons
with a band of uncorrelated conduction electrons. These two electronic
sub-systems are coupled by a hybridization term. The Hamiltonian of
the PAM reads:
\begin{align}
  \label{hamiltonian}
    H =&\sum_{\vec{k},\sigma}
    \epsilon(\vec{k}) s_{\vec{k}\sigma}^{\dagger}s_{\vec{k}\sigma} + 
    \sum_{i,\sigma} \epsilon_{\rm f} f_{i\sigma}^{\dagger}f_{i\sigma} +
    V \sum_{i,\sigma} (f_{i\sigma}^{\dagger}s_{i\sigma} +
    s_{i\sigma}^{\dagger}f_{i\sigma} ) +\\
    & +\frac{1}{2} U \sum_{i,\sigma}
    n^{\rm (f)}_{i\sigma}n^{\rm (f)}_{i-\sigma}
    - \sum_{i\sigma} z_{\sigma} b_{\rm ext}
    (n^{\rm (f)}_{i\sigma}+n^{\rm (s)}_{i\sigma})\nonumber
\end{align}
Here, $s_{\vec{k}\sigma}$ ($f_{i\sigma}$) and
$s_{\vec{k}\sigma}^{\dagger}$ ($f_{i\sigma}^{\dagger}$) are the
creation and annihilation operators for a conduction electron with
Bloch vector $\vec{k}$ and spin $\sigma$ (a localized electron on site
$i$ and spin $\sigma$) and
$n_{i\sigma}^{\rm (f)}=f_{i\sigma}^{\dagger}f_{i\sigma}$
($s_{\vec{k}\sigma}=\frac{1}{N}\sum_{\vec{k}} e^{i \vec{k} \vec{R}_i}
s_{i\sigma}$).
The dispersion of the 
conduction band is $\epsilon(\vec{k})$ and $\epsilon_{\rm f}$ is the
position of the localized level.
The hybridization strength $V$ is taken to be
$\vec{k}$-independent, and finally $U$ is the on-site Coulomb
interaction strength between two $f$-electrons. Throughout this paper, the
conduction band will be described by a free (Bloch) density of states,
$\rho_0(E)= \frac{1}{N}\sum_{\vec{k}} \delta (E-\epsilon(\vec{k}))$,
of semi-elliptic shape. Its width $W=1$ sets the
energy scale, and its center of gravity the energy-zero: 
$T_{ii}=\frac{1}{N}\sum_{\vec{k}}\epsilon(\vec k)\stackrel{!}{=}0$.
The magnetic field $b_{\rm ext}$ is given in energy units of
the band width $W$ (
$z_{\sigma}=+ 1 (-1)$ for $\sigma=\uparrow(\downarrow)$). The external field couples equally
to the spin of the $f$ and conduction band electrons.
This is in our opinion most
appropriate for the model
Hamiltonian~(\ref{hamiltonian}) where the $f$-states are taken to be
non-degenerate ($s$-type).
Without using the full orbital degeneracy, one could alternatively use the g-factors
of the real materials. Other authors have even completely
neglected the coupling of the magnetic field to conduction band states
arguing that the corresponding $g$ factor is negligible\cite{SI96,Ono96,Ono98}.

We employ the dynamical mean-field
theory (DMFT)\cite{GKKR96} in combination with the modified perturbation
theory (MPT)\cite{MWPN99} to determine the one-electron Green's function, from
which the
excitation spectrum as well as magnetization, effective mass and other
quantities can be calculated.
This method has previously been applied to the
paramagnetic\cite{MN00b} and the ferromagnetic PAM\cite{MN00c},
so we can confine ourselves to a short summary: The underlying idea of
the DMFT is that the local self-energy such as occurs in the limit of infinite
spatial dimensions\cite{MV89,Mue89}, can be taken to be
that of an appropriately defined single-impurity Anderson
model. We solve the latter using the \textit{modified perturbation theory}.
Our starting point is the following ansatz for the
self-energy\cite{MFBP82,KK96}:
\begin{equation}
  \label{eq:ansatz}
  \Sigma_{\sigma}(E)=U \langle n_{-\sigma}^{(f)}\rangle
  +\frac{\alpha_{\sigma} \Sigma_{\sigma}^{\rm (SOC)}(E)}
  {1-\beta_{\sigma} \Sigma_{\sigma}^{\rm (SOC)}(E)}
\end{equation}
$\alpha_\sigma$ and $\beta_{\sigma}$ are introduced as parameters to be
determined
later. $\Sigma_{\sigma}^{\rm (SOC)}(E)$ is the
second-order contribution to perturbation theory around the Hartree-Fock 
solution\cite{Yam75}. 
Equation~(\ref{eq:ansatz}) can be understood as the simplest possible
ansatz which can, on the one hand, reproduce the perturbational result
in the limit $U\rightarrow 0$, and, on the other hand, recovers the
atomic limit for
appropriately chosen $\alpha_{\sigma}$ and $\beta_{\sigma}$\cite{MFBP82}.

Using the perturbation theory around the Hartree-Fock solution
introduces an ambiguity into the calculation. Within the self-consistent
Hartree-Fock calculation, one can either choose the chemical potential
to be equivalent to the chemical potential of the full MPT calculation,
or take it as parameter $\tilde{\mu}$ to be fitted to another physically
motivated constraint.
In reference~\onlinecite{KK96} the
Luttinger theorem\cite{LW60}, or equivalently the Friedel sum
rule\cite{Fri56,Lan66}, was used to determine $\tilde{\mu}$.
As discussed in  Ref.~\onlinecite{MN00c},
we use the physically motivated condition of
identical impurity occupation numbers for the Hartree-Fock and 
the full calculation ($n_{\sigma}^{(f,{\rm HF})}=n_{\sigma}^{(f)}$) to
determine $\tilde{\mu}$, which also allows for a consistent extension of the
method to finite temperatures\cite{MN00b,MN00c}. Except for symmetric
parameters this will lead to an approximate
fulfillment of the Luttinger theorem only\cite{MWPN99}. The key features
of the results presented
below, however, are not decisively influenced by this shortcoming.
This was checked by adopting the alternative approach of fitting
$\tilde{\mu}$ using the Friedel sum rule.
A more detailed analysis of the
different possibilities to determine $\tilde{\mu}$ is found in
reference~\onlinecite{PWN97} where the DMFT-MPT was applied to the single-band
Hubbard model.
Finally, the parameters $\alpha_{\sigma}$ and $\beta_{\sigma}$
have to be determined. Instead of using the ``atomic'' limit of $V=0$ as 
was done for example in references~\onlinecite{MFBP82,MLFT86,KK96}, we
make use of the moments of the spectral density. This procedure is
described in detail in references~\onlinecite{PWN97,MWPN99}. The result
not only
fulfills the $V=0$ limit, but also recovers the high-energy behavior of
the Green's function up to the order $(\frac{1}{E^4})$.

It has been shown that the approximation
scheme, as described above, gives qualitatively reliable results by
comparing with
numerical renormalization group theory and quantum Monte Carlo
calculations for two different ``strong-coupling effects'', namely the
Mott-Hubbard insulator\cite{GKKR96,BCV00pre} and ferromagnetism in the
single-band Hubbard model\cite{Ulm98,PHWN98}. 
The periodic Anderson model was investigated by this and related
methods in the paramagnetic\cite{VTJK00,MN00b} and the ferromagnetic
phase\cite{MN00c,MN00d}. In the following we present results for the
periodic Anderson model in the paramagnetic, but close to the
ferromagnetic region of the phase diagram\cite{MN00c}.

\begin{figure}[t]
    \includegraphics[width=7cm]{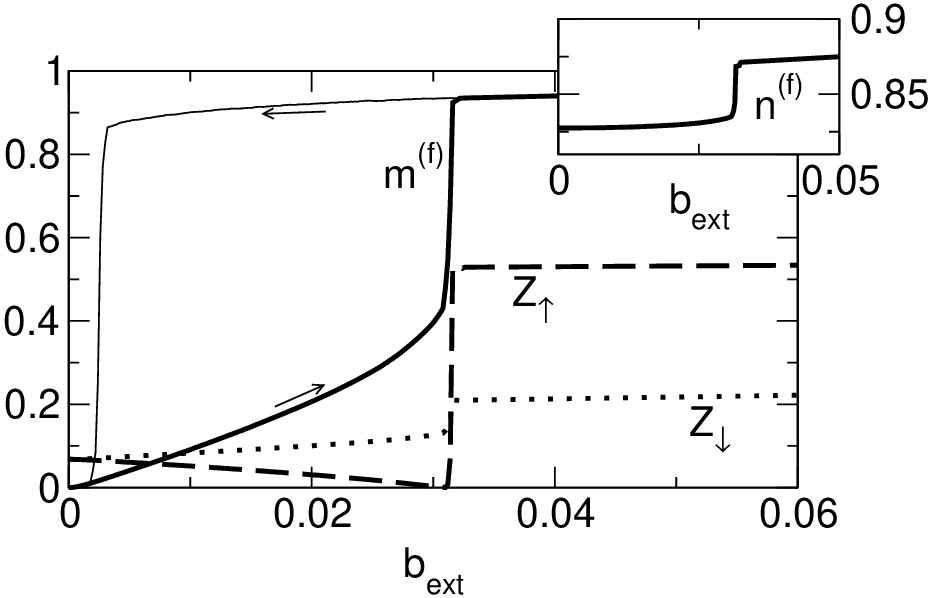}
    \caption{$m^{\rm (f)}(b_{\rm ext})$ (solid line) and the (spin-dependent)
      quasiparticle weight $Z_{\sigma}$ as function of $b_{\rm ext}$ for
      $\epsilon_{\rm f}=-0.4$, $U=4$, $n^{\rm (tot)}=1.85$, $V=0.2$ and
      $T=0$. The inset shows $n^{(\rm (f)}$ as function of
      $b_{\rm ext}$.}
    \label{fig:m_B_ef-0.4}
\end{figure}

In Fig.~\ref{fig:m_B_ef-0.4}, the f-magnetization $m^{\rm (f)}=
\frac{n^{\rm (f)}_\uparrow - n^{\rm (f)}_\downarrow} {n^{\rm
    (f)}_\uparrow + n^{\rm (f)}_\downarrow}$
and the (spin-dependent)
quasiparticle weight $Z_{\sigma}=(m^*)^{-1}=(1-\frac{\partial
  \Sigma_{\sigma}(E)}{\partial E}|_{E=0})^{-1} $ 
are plotted as function of the external field $b_{\rm ext}$ for 
$\epsilon_{\rm f}=-0.4$, $U=4$, $n^{\rm (tot)}=1.85$, $V=0.2$ and zero
temperature.  These parameters away from the symmetric point (where
$\epsilon_{\rm f}=-\frac{U}{2}$) are close to the ferromagnetic phase which
would be reached for lower electron density ($n^{\rm (tot)}\lesssim
1.7$)\cite{MN00c}. The increase of $m(b_{\rm ext})$ (thick solid line)
shows a sharp rise at $b_{\rm ext}=b_{\rm ext}^*\approx 0.033$. We call this
phenomenon a metamagnetic transition. 
In Fig.~\ref{fig:m_B_ef-0.4}, a discontinuous jump in $m^{\rm (f)}$ is
visible. However, due to the finite numerical resolution of our
calculations we cannot rule out a continuous transition.
For a range of $b_{\rm ext}<b_{\rm ext}^*$ a
high-magnetization and a low-magnetization self-consistent solution
co-exist. For $T=0$, the
latter is always stable. The spin-$\uparrow$
quasiparticle weight vanishes at $b_{\rm ext}^*$ implying a maximum in the
effective mass. For $b_{\rm ext}>b_{\rm ext}^*$, the quasiparticle weight
is significantly larger than for $b_{\rm ext}=0$.
As shown in the inset, the
metamagnetic transition is also accompanied by a sharp increase in
$f$-occupancy $n^{\rm (f)}$ at $b_{\rm ext}=b_{\rm ext}^*$.

\begin{figure}[t]
    \includegraphics[width=7cm]{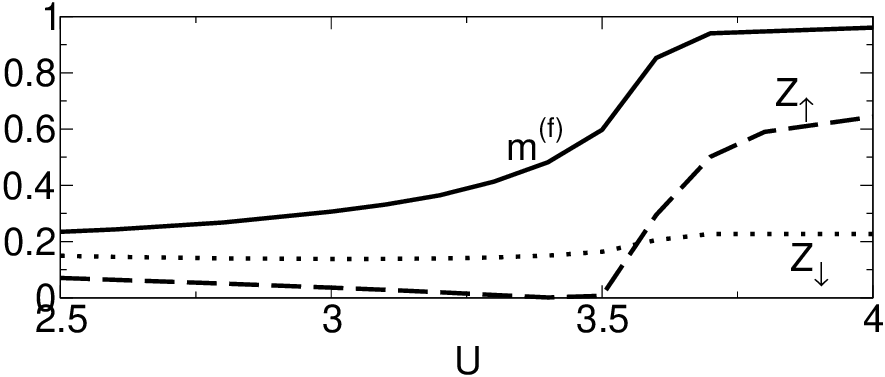}
    \caption{Same as Fig.~\ref{fig:m_B_ef-0.4} for $b_{\rm ext}=0.1$ as
      function of $U$.}
    \label{fig:m_U_bext}
\end{figure}
In Fig.~\ref{fig:m_U_bext}, the magnetization and the quasiparticle
weights are shown as function of interaction strength $U$ for fixed
$b_{\rm ext}=0.1$. At $U=U^*$ a metamagnetic transition is also observable,
being accompanied by the same behavior of
$Z_{\sigma}$ as discussed above.

\begin{figure}[t]
    \includegraphics[width=7.5cm]{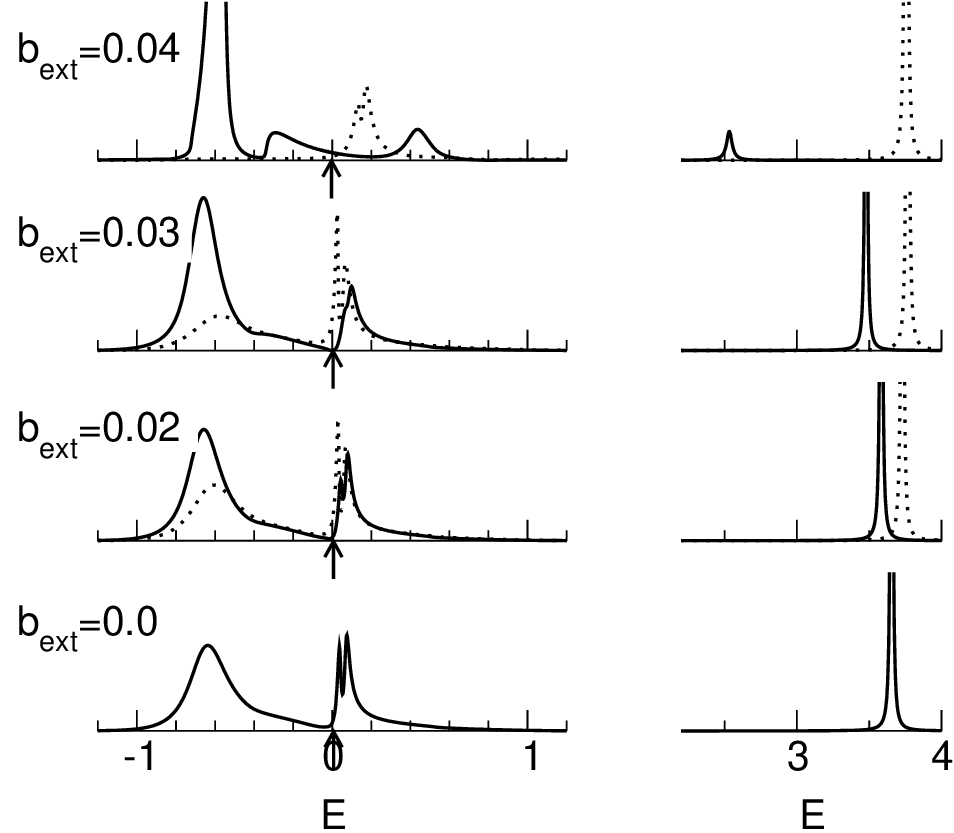}
    \caption{$f$-DOS for the parameters of Fig.~\ref{fig:m_B_ef-0.4}
      with $b_{\rm ext}$ as indicated in the different rows. The solid (dotted) line
      corresponds to the spin-$\uparrow$ ($\downarrow$) channel. The
      two columns show the same DOS at different energy ranges: lower
      charge excitation and Kondo resonance in the left
      and the upper charge
      excitation in the right column. The chemical
      potential is given by the arrows.}
    \label{fig:dos_t0_extfield}
\end{figure}
The $f$-DOS for $b_{\rm ext}\in\{0.0, 0.02, 0.04, 0.06\}$ is plotted in
Fig.~\ref{fig:dos_t0_extfield} for the same parameters used in
Fig.~\ref{fig:m_B_ef-0.4}. For zero-field, the DOS
consists of lower
charge excitation at $\epsilon_{\rm f}$, upper charge excitation at $\epsilon_{\rm f}+U$ and the
Kondo resonance at $E\approx\mu$ which is split by the coherence
gap\cite{MN00b}. Applying
a small external magnetic field ($b_{\rm ext}\leq0.04$) induces a Zeeman
shift proportional to
$b_{\rm ext}$, which is best
visible in the upper charge excitation. The Kondo resonances in the
spin-$\uparrow$ and $\downarrow$ channel seem also to be shifted, however, in
the opposite direction: Spin-$\uparrow$ to higher, and spin-$\downarrow$
DOS to lower energies (``inverse Zeeman shift'').
This is very unexpected behavior and is only found for parameters
that show a metamagnetic transition. For other parameters further away
from the ferromagnetic phase (e.g.\ $U<U^*$ or also at the symmetric
point $\epsilon_{\rm f}=-\frac{U}{2}$)
a ``normal'' Zeeman shift is observed for the Kondo
resonance\cite{MeyerDiss}.
Closer investigation shows that the apparent shift is indeed no shift,
but a suppression of spectral weight in the spin-$\uparrow$ channel just
below the Fermi energy. This suppression is also indicated by a
decrease of $Z_{\uparrow}$ as seen in Fig.~\ref{fig:m_B_ef-0.4}.

For fields above the metamagnetic transition, the
picture changes dramatically. The lower charge excitation is fully
saturated and the shift between the two spin channels of the upper
charge excitation
is much larger than one would expect from an extrapolation from the other
figures. Finally, the Kondo resonance
disappears in the spin-$\uparrow$ DOS. A broad structure 
located well above the chemical
potential is visible in the
$\downarrow$-channel only. These DOS strongly resemble those of the
ferromagnetic PAM\cite{MN00c} whereas the DOS for lower fields
more or less correspond to the paramagnetic
DOS\cite{MN00b} plus the Zeeman shift. So the
metamagnetic transition closely resembles the para- to ferromagnetic
transition\cite{MN00c}. This agrees with the fact
that the metamagnetic transition is clearly a strong coupling phenomenon
as was also found for ferromagnetism in the intermediate-valence
regime\cite{MN00d}
(see Fig.~\ref{fig:m_U_bext}).

Let us shortly discuss the nature of the metamagnetic transition
exhibited by our results. The origin of the transition
is somewhat different from previously discussed approaches to
metamagnetism, such as the Kondo-volume collapse
approach\cite{AM82,Ohk89,SO98}, or explanations based on special
features of the density of
states\cite{Bre87,Aea98,SO98,Ono96,Ono98}. For example, solving the PAM
within self-consistent perturbation theory\cite{CQS93,Mut00}, one also finds a
metamagnetic transition. Here, however, the transition originates from sharp features
in the excitation spectrum which are already present for zero-field. On
applying the field, the
Zeeman shift can push these features across the chemical potential and
thus dramatically change the spin-dependent f-occupation and hence the
magnetization. In our results however, the transition
can be traced back to dramatic changes of the correlation-induced
features of the excitation spectrum due to the external magnetic field.
This clearly distinguishes the scenario presented in this paper from
previously discussed metamagnetic transitions in the PAM. 
This manifests itself furthermore in the observed hysteresis which cannot
occur in a scenario where the transition is due to the Zeeman shift of a
strongly peaked DOS.

A phenomenologically similar metamagnetic transition is known to exist
for the half-filled Hubbard model\cite{LGK94,GKKR96,JC00}. Here, a
jump in the magnetization as function of external field is found for
interaction strengths $U$ close to the critical $U_c$ separating the
Mott-insulator from the metallic regime. Similar to the transition
described in this paper, the transition in the Hubbard model also shows a
hysteresis and is indicated by a suppression of the Kondo resonance.
Another similarity
might be seen in the fact that in both metamagnetic transitions, the
high-magnetization state is characterized by a stronger localization of
the correlated electrons.
In the Hubbard model, this transition is found for
half-filling where it leads from a metallic to an insulating state. 
In the transition presented here we also see a tendency towards stronger localization of the
f-electrons. Contrary to the situation in the half-filled
Hubbard model, however, the PAM is metallic above and below the critical
field. Furthermore, the suppression of the Kondo resonance occurs
only in the proximity of the ferromagnetic phase, and only in the
spin-$\uparrow$ channel.

The metamagnetic transition is not simply an enlargement of the
'ferromagnetic' phase caused by the magnetic field. If this were the case one would expect 
similar behavior using
other approximation methods, which yield almost the
same ferromagnetic phase diagram as the MPT\cite{MN00d} such as the
modified alloy 
analogy and the spectral density approximation for the
PAM\cite{MN00d,RMSRN00,MNRR98}. However, both methods do not show the
metamagnetic
transition. The low-energy (``Kondo'') physics, which are not recovered
by the other methods\cite{MN00d} play a decisive role.

What is the relevance of our results to the metamagnetic
transitions observed experimentally? The most well-known example of a
heavy-fermion compound showing a metamagnetic transition from para- to a
``ferromagnetic'' state is
CeRu$_2$Si$_2$\cite{Aea98,RERLF88,LVHLF89}. This material
exhibits peculiar behavior alongside the metamagnetic
transition. The effective mass shows a sharp maximum at $b_{\rm
  ext}=b_{\rm ext}^*$, and for $b_{\rm ext}>b_{\rm ext}^*$ is
suppressed\cite{AUAO93,Mea91,Sea95,Aea98}, which is exactly the behavior of
the metamagnetic transition of the PAM discussed above. Furthermore, the
experimentally confirmed\cite{AUAO93}
stronger localization of the $f$-electrons in the high-magnetization
state is indicated by the 
jump in $n^{\rm (f)}$ that we discussed above..

There are two findings for CeRu$_2$Si$_2$, which are
not reproduced by our approximation.
Neutron scattering experiments\cite{RERLF88}
revealed antiferromagnetic inter-site correlations for $b_{\rm
  ext}<b_{\rm ext}^*$. These
correlations can not be found within our approximation
scheme since it is based on dynamical mean-field theory.
However, the parameters we have used above are 
close to the antiferromagnetic regime of the PAM\cite{TJF97}. It seems
therefore reasonable to expect
antiferromagnetic correlations in the zero-field (low-magnetization)
state. The second effect is the volume effect of the
metamagnetic transition\cite{LVHLF89,LVPLHFVO89}. A Kondo-volume collapse similar to
that discussed in connection with the $\gamma\rightarrow\alpha$
transition of Ce\cite{AM82} has been put forward as possible source of the
metamagnetic transition in CeRu$_2$Si$_2$\cite{Ohk89,SO98}. 
Our model does not include any coupling to the lattice. However, the
significant change in $f$-occupancy that we find should result in a
volume change, if the lattice coupling were to be included. This could
even amplify the transition. 

Contrary to our results, the metamagnetic
transition in CeRu$_2$Si$_2$
does not show hysteresis\cite{Sea95}. Let us point out that in our
results we find merely a ``mathematical hysteresis''. In
Fig~\ref{fig:m_B_ef-0.4}, the thick line is the physically
meaningful result whereas the thin line represents only a mathematical
solution to the equations to which we cannot ascribe a clear physical
meaning. However, one could assume that for finite temperatures, a true
hysteresis would be found. Experimentally, a metamagnetic transition
from a paramagnetic state that is accompanied by a hysteresis is
realized in UCoAl\cite{MGKYASIS99}. This material, however, does not
show a sharp maximum in the effective mass of the
quasiparticles\cite{Aea98} so it is unclear whether the picture
described in this paper is relevant for this material.

At this point, let us comment
on the critical fields necessary to drive the metamagnetic transition.
In our units, $b_{\rm ext}=0.033$ corresponds to several hundreds of
Tesla if the bandwidth would be of order $1 eV$. This is much
too large compared to the above-cited experimental results. We believe
this is a limitation of the MPT approximation. In our explanation, the
transition
is closely connected with the low-energy properties, the existence
and the width of the Kondo resonance. The MPT, however, tends to
overestimate the low-energy scales\cite{MN00b}, so the absolute
values of the critical field as obtained by our calculation should not
be taken too seriously.

To summarize, we have presented a new scenario for a metamagnetic
transition based on a strong electron-electron
coupling effect. For a periodic Anderson model, with parameters located
in the paramagnetic regime between the
ferro- and the antiferromagnetic phase, an increase of the external
field for sufficiently high interaction
strength leads to a sudden sharp increase of the
magnetization (metamagnetic transition). This is accompanied by
a suppression of low-energy spectral
weight (Kondo resonance) in the spin-$\uparrow$ DOS.
The features of the transition show strong similarities
to the experimentally observed metamagnetic transition of
CeRu$_2$Si$_2$. 
The decreasing spectral weight near the Fermi level
prior to the metamagnetic transition could serve as an experimentally
accessible indicator for the relevance of the
proposed mechanism to the real physics of CeRu$_2$Si$_2$.
\acknowledgements
We want to thank A. Hewson and Y. Ono for pleasant and helpful
discussions.

\end{document}